\documentclass[aps,twocolumn,superscriptaddress,floatfix,longbibliography]{revtex4-1}
\usepackage{tikz}
\usepackage{lipsum}
\usepackage{graphicx}
\usepackage[export]{adjustbox}
\usepackage{amsmath,amssymb,wasysym}
\usepackage{braket}
\usepackage{mathtools}
\usepackage{mathrsfs}
\usepackage{verbatim}
\usepackage{makecell}
\usepackage{cases}
\usepackage{bbm}
\usepackage{hyperref}

\newcommand{\Z}{\mathbb{Z}}

\newcommand{\abs}[1]{\left| #1\right|}

\DeclareMathOperator{\Tr}{Tr}

\newcommand{\lOTOC}{\lambda_{\text{OTOC}}}
\newcommand{\lclas}{\lambda_{\text{chaos}}}
\newcommand{\lsaddle}{\lambda_{\text{saddle}}}

\newcommand{\heff}{\hbar_{\text{eff}}}

\newcommand{\fig}[1]{Fig.~\ref{fig:#1}}

\begin{document}
\title{Does scrambling equal chaos?}
\author{Tianrui Xu}
\affiliation{Department of Physics, University of California, Berkeley, CA 94720, USA}
	\affiliation{Materials Sciences Division, Lawrence Berkeley National Laboratory, Berkeley, California 94720, USA}
\author{Thomas Scaffidi}
	\affiliation{Department of Physics, University of Toronto, Toronto, Ontario, M5S 1A7, Canada}
	\author{Xiangyu Cao}
\affiliation{Department of Physics, University of California, Berkeley, CA 94720, USA}
	\begin{abstract}
 Focusing on semiclassical systems, we show that the parametrically long exponential growth of out-of-time order correlators (OTOCs), also known as scrambling, does not necessitate chaos. Indeed, scrambling can simply result from the presence of unstable fixed points in phase space, even in a classically integrable model. We derive a lower bound on the OTOC Lyapunov exponent which depends only on local properties of such fixed points. We present several models for which this bound is tight, i.e. for which scrambling is dominated by the local dynamics around the fixed points. We propose that the notion of scrambling be distinguished from that of chaos.
\end{abstract}
\maketitle

\textit{Introduction.}-- Classical chaos is a ubiquitous phenomenon in nature. It explains how a deterministic dynamical system can be inherently unpredictable due to exponential sensitivity to initial conditions (the butterfly effect), and is a foundation of thermodynamics and of hydrodynamics. By contrast, the notion of ``quantum chaos'' is not as sharply defined, and carries multiple meanings resulting from several waves of attempts to extend the notion of chaos into the quantum world.
Forty years ago, several groups of authors~\cite{casati,bohigas,berry-tabor,BERRY1981163} famously pointed out that the quantization of classical systems leaves a footprint in the level statistics of the energy spectrum.  Since then, the dichotomy of random-matrix \textit{vs} Poisson level statistics has become a standard diagnostic of quantum integrability, the lack of which is considered by many as a definition of quantum chaos. Several other diagnostics have been considered ever since, including Loschmidt echo~\cite{echo}, dynamical entropy~\cite{connes1987,AF}, decoherence~\cite{zurek}, entanglement~\cite{Nie_2019,dubail,chaos_ent}, etc, forming a large ``web of diagnostics''~\cite{nie_web}.

Recently, progress in the study of quantum information, black holes, and holography~\cite{Hayden_2007,Sekino_2008,Lashkari:2011yi,BHbutterfly,Shenker:2014cwa, Yoshida:2017non,yoshidayao,monroe_exp} has led to yet another putative definition of quantum chaos (which we shall refer to as  ``scrambling'', following Sekino and Susskind~\cite{Sekino_2008}), in terms of out-of-time order correlators (OTOCs). Its definition~\cite{larkin,Maldacena:2015waa} is directly motivated by the butterfly effect. More precisely, one starts from the observation that the sensitivity to initial condition can be quantified by a Poisson bracket:  $\{q(t), p\} = \partial q(t)/ \partial q(0)$ where $q$ and $p$ are a conjugate pair. The OTOC is then defined as the thermal average of the square of a commutator $[\hat{q}(t), \hat{p}]$, by ``quantizing'' $\{q(t), p\}$.

The behavior of OTOCs has been studied in a wide range of quantum systems, and they turn out to be most useful in characterizing large-$N$ systems dual to semi-classical gravity via the holographic principle. In such systems, the OTOCs can have exponential growth, which has been interpreted as a signature of quantum chaos ever since \cite{Maldacena:2015waa}. The growth rate is referred to as a (quantum) Lyapunov exponent, and bounds thereof are called ``bounds on chaos'' \cite{Maldacena:2015waa,pcasa18,bound_srednicki,Avdoshkin}. ``Maximally chaotic'' systems, which saturate those bounds, received particular attention as canonical toy models of strongly coupled systems and of holography~\cite{Sachdev:1992fk,Kitaev:2015,Maldacena:2016hyu,Sachdev:2015efa,kitaev17}. 

\begin{figure}
    \centering
    \includegraphics[width=.95\columnwidth]{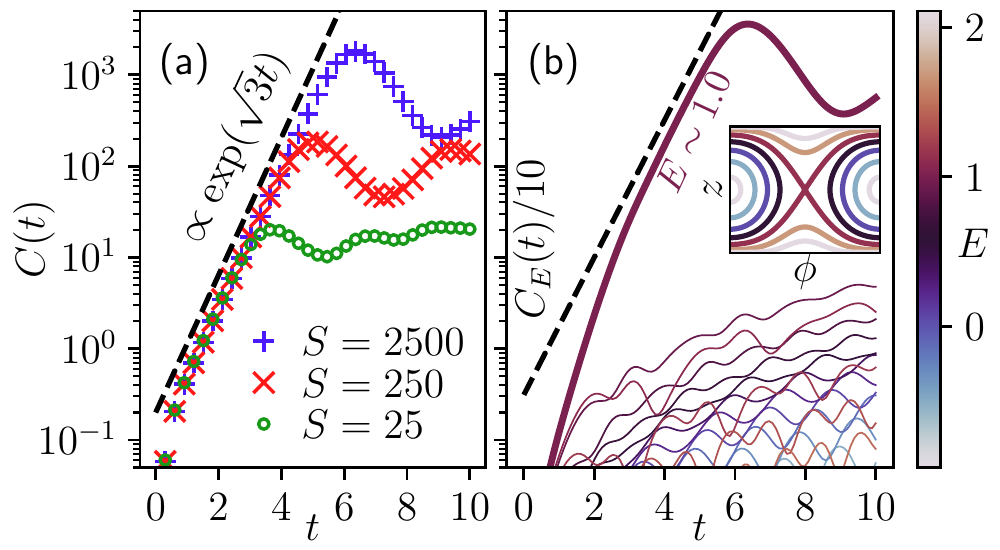}
    \caption{(a) Extended exponential growth of the infinite-temperature OTOC~\eqref{eq:OTOC} of the integrable LMG model \eqref{eq:H_classical} in the semi-classical limit. The growth saturates at the Ehrenfest time $\sim \ln (S)$. The exponent $\lOTOC = \sqrt{3}$ is the unstable exponent of the saddle point in the classical phase space. (b) Microcanonical-ensemble OTOCs $C_E(t)=-\frac1{125}\sum_{\epsilon \in b_E}\braket{\epsilon|[\hat{S}_z, \hat{S}_z(t)]^2|\epsilon}$ ($S = 2500$), where $b_E$ is an energy window of $125$ Hamiltonian eigenstates $\{\vert \epsilon \rangle \}$ with average energy $E$. A few representative ensembles across the entire energy spectrum are shown. The one with $E \approx 1$, corresponding to the classical saddle, dominates the exponential growth observed in (a). \textit{Inset}: Energy landscape of the classical limit, with the same color code as (b), and the saddle in the center.} 
    \label{fig:OTOC}
\end{figure}
Nevertheless, the interpretation of exponential OTOC growth as chaos is questionable, especially in the context of quantum systems in the semiclassical limit. (Refs~\cite{sarang1,sarang2} discussed the issue far from classical limit.) There, ``chaos'' has an unambiguous meaning: the distance between a typical pair of neighboring trajectories grows exponentially in time. The standard quantitative measure of chaos is the maximal Lyapunov exponent, $\lclas$, defined by the \emph{phase space average of the log} of sensitivity~\cite{Politi:2013}. This differs from the exponential growth rate of an OTOC, $\lOTOC$, which is rather the \emph{log of the phase space average} of sensitivity squared. Since the log of the average is larger than the average of the log, we have~\cite{galitski} (see also \cite{cotleraop18,Politi:2013,dicke,rozenbaum19,pilatowsky2019positive,rey}):
\begin{equation}
    \lOTOC \ge 2 \lclas \,. \label{eq:bound1}
\end{equation} 
Presented as such, the difference between scrambling and chaos might seem an innocuous quantitative detail, and is often so considered. In this work, we argue that, to the contrary, the difference is \textit{qualitative}: scrambling can occur independently of chaos. We shall identify one simple alternative mechanism: isolated saddle points. Indeed, the unstable trajectories in a small neighborhood of a saddle can be enough for the OTOC to grow exponentially. Such contributions lead to another bound:
\begin{equation}
    \lOTOC \ge \lsaddle \,,\label{eq:bound2}
\end{equation}
where $\lsaddle$ can be simply calculated in terms of the local properties of the saddle, see below. As a result, OTOCs can grow exponentially in non-chaotic systems. Furthermore, even in chaotic systems, scrambling can be dominated by saddles instead of chaos, i.e., $\lsaddle$ is closer to $\lOTOC$ than $\lclas$. These findings suggest that scrambling and chaos should better be treated as distinct concepts We note that Ref.~\cite{rozenbaum19} made a similar case using a distinct argument, and that the lack of a clear distinction between integrable and non-integrable behavior for OTOCs was also found away from the semiclassical limit in Refs~\cite{sarang1,sarang2}.

\textit{Two-dimensional case.}-- 
As an illustrative example, we consider a 
special instance of the Lipkin-Meshkov-Glick (LMG) model~\cite{LIPKIN1965188,*GLICK1965211,*MESHKOV1965199,pilatowsky2019positive,excited}, which is integrable. In the classical limit, it is defined by the Hamiltonian
\begin{equation}
    H = x + 2 z^2 \,.  \label{eq:H_classical} 
\end{equation}
where $x, y, z$ form a classical SU(2) spin satisfying $x^2+y^2+z^2 = 1$ and $\{x, y\} = z$, etc. It is easy to check that $(x,y,z) = (1, 0,0)$ is a saddle point. Linearizing the dynamics close to it leads to local coordinates $a_{\pm}$ satisfying equations of motion 
\begin{equation}
\frac{d a_\pm}{dt } \approx \pm \omega a_{\pm} \,,\, \omega = \sqrt{3} \,, \label{eq:EOM} \end{equation}
near the saddle.  Of course, such a fixed point is \textit{not} considered chaotic~\cite{Politi:2013}, since $a_+$ grows only exponentially near the saddle.

We now compute an OTOC in the quantization of \eqref{eq:H_classical}. Namely, we consider the quantum Hamiltonian $\hat{H} = \hat{x} + 2 \hat{z}^2$ where $\hat{x},\hat{y}, \hat{z} = \hat{S}_x / S,  \hat{S}_y / S, \hat{S}_z / S$ are rescaled $SU(2)$ spin operators with spin $S$. They satisfy the commutation relations such as $[\hat{x}, \hat{y}] = i \heff \hat{z}$, where $\heff = 1/S$ is the effective Planck constant ($\heff \to 0$ is the classical limit)~\footnote{Whenever possible, we prefer semiclassical OTOCs to classical ones in numerics, to avoid bias and statistical noise in phase space averaging.}. The OTOC is defined at infinite temperature, with respect to the operator $ \hat{O}= \hat{z}$:
\begin{equation}
     C(t) := \frac1{\heff^2} \frac{\Tr\left([\hat{O}(t),\hat{O}]^\dagger[\hat{O}(t),\hat{O}]\right)}{ \Tr(\mathbbm{1})} \,.\label{eq:OTOC}
\end{equation}
The numerical result, Fig.~\ref{fig:OTOC}, shows an extended period of exponential growth, up to the Ehrenfest time
\begin{equation}
    C(t) \sim e^{\lOTOC t} \,,\, 1 \lesssim  t \lesssim  \ln (1/\heff) \,,
\end{equation}
with the Lyapunov exponent $\lOTOC = \omega = \sqrt{3}$ precisely.

To explain this observation, let us focus on the classical limit. Then, the OTOC \eqref{eq:OTOC}, which is an infinite-temperature average of a commutator squared, becomes the following phase space average of sensitivities squared~\cite{cotleraop18} :
\begin{equation}
    C(t) =  \int_{\mathbb{S}^2} \abs{\{z(t), z \}}^2 d \mathcal{A}  = \int_{\mathbb{S}^2}  \abs{\frac{\partial z(t)}{\partial \phi}}^2 d \mathcal{A} \,,
\end{equation}
where $d \mathcal{A}$ is the normalized area form on the sphere $\mathbb{S}^2$ and $\phi$ is the azimuthal angle and conjugate to $z$. The integrand is not exponentially growing in $t$, \textit{except} near the saddle point. Indeed, in a narrow strip $$ \mathcal{S}_t = \{|a_+| < \delta e^{-\omega t}, |a_-| < \delta \} \subset \mathbb{S}^2 \,, $$ of volume $\delta^2  e^{-\omega t}$~\footnote{$\delta$ is some small constant that does not depend on $t$; in other words, $\mathcal{S}_t$ is a neighborhood of the unstable invariant manifold  $\{ a_{+} = 0 \}$ near the fixed point.}, the linearized dynamics \eqref{eq:EOM} is a valid approximation up to $t$, until which point the sensitivity grows exponentially: $\left|\frac{\partial z(t)}{\partial \phi}\right|  \sim e^{ \omega t}$. Now, recall that the OTOC involves the \textit{square} of the sensitivity, which overwhelms the exponentially small volume. So, $\mathcal{S}_t$ alone contributes an exponential growth:
\begin{equation}
    C(t) \ge \int_{\mathcal{S}_t} \abs{\frac{\partial z(t)}{\partial \phi} }^2 d \mathcal{A}  \sim e^{2 \omega t} \times   \delta^2  e^{-\omega t} = \delta^2 e^{\omega t} \,. \label{eq:Ct_strip}
\end{equation}
This leads to the following lower bound on $\lOTOC$:
\begin{equation}
    \lOTOC \ge \lsaddle := \omega \,. \label{eq:bound_ex}
\end{equation} 
In the case of the LMG model, this bound is tight because the saddle point is the \textit{only} source of scrambling. Indeed, the OTOC in a microcanonical ensemble has significant growth only for energies close to that of the saddle point, see Fig.~\ref{fig:OTOC}(b). 

We have thus demonstrated by a simple example that OTOCs can grow exponentially in a classical integrable system which has a saddle point. This principle applies to any saddle points in a two-dimensional phase space. We remark that the analysis here is distinct from earlier works~\cite{pappalardi18,hummel19,pilatowsky2019positive,rozenbaum19}. Some of them suggested a bound $\lOTOC \ge 2 \omega$, which differs from \eqref{eq:bound_ex} by the small volume factor. In the most recent  \cite{pilatowsky2019positive}, this difference results from using a variant of OTOC involving an initial wave-packet localized at the saddle, making an exponential spreading rather expected. In contrast, our point here is that an attempt to diagnose chaos in a finite portion of phase space (using an OTOC with ensemble average) can be failed by false positives.        

\textit{General case.}-- The above reasoning can be directly generalized to a fixed point in an $n$-dimensional phase space. Again, we linearize the dynamics near it, that is, we can find some local (complex) coordinate system $(x_1, \dots, x_{n})$, such that $\dot{x_j} = (\omega_j + i \eta_j) x_j$, where 
$$ \omega_1 \ge \dots \ge \omega_m > 0 \ge \omega_{m+1} \ge \dots \ge \omega_{n}  $$ 
are the real part of stability exponents, and $m$ is the number of unstable ones~\footnote{This is the generic,  diagonalizable, case. Otherwise there are extra terms due to nontrivial Jordan blocks, which will only lead to polynomial corrections to the exponential growth.}. 
Then consider the ``hypercuboid'' defined as
$$ 
\mathcal{S}_t = \{ |x_i|  < \delta e^{-\omega_i t} \ \forall \ i \le m, |x_i|  < \delta  \ \forall \ i > m \}
$$
It has volume $\mathrm{Vol}( \mathcal{S}_t)  \sim \delta^{2n} e^{-\sum_{j \le m} \omega_j t}$, and almost any initial condition within it has exponentially growing sensitivity $\sim e^{\omega_1 t}$ up to time $t$.  It follows, by a similar calculation as Eq.~\eqref{eq:Ct_strip} above, that the localized contribution from $\mathcal{S}_t$ leads to a lower bound on the OTOC Lyapunov exponent:
\begin{equation}
    \lOTOC \ge   \lsaddle :=   \omega_1 - \sum_{j>1, \omega_{j} >0} \omega_j \,.  \label{eq:lambdaL_gen}
\end{equation}
This bound is a generalization of \eqref{eq:bound_ex}, and reduces to it when there is a single unstable exponent. The bound in Eq.~\eqref{eq:lambdaL_gen} is of course nontrivial only if $\lsaddle > 0$. We will give several examples below where that is the case.



\begin{figure}
    \centering
    \includegraphics[width=.8\columnwidth]{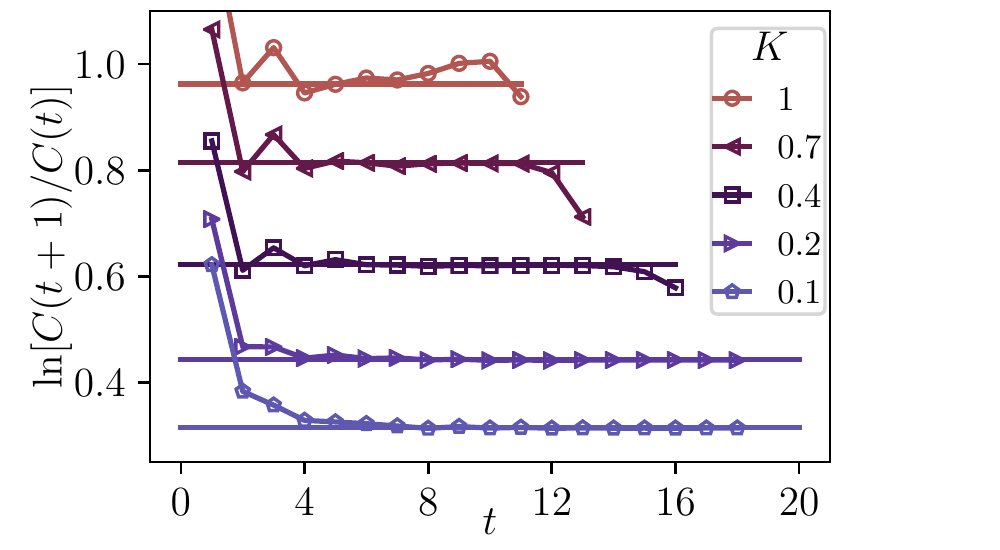}
    \caption{The markers show the instantaneous exponential growth rate $\ln [C(t+1)/C(t)]$ as a function of the number of kicks $t$ in the kicked rotor model, quantized with Planck constant $\heff = 2^{-14}$ (see Ref.~\cite{galitski} for definition and methods). Good agreement can be seen with the saddle point exponents $\omega(K)$ from Eq. \eqref{eq:omega_rotor}, plotted as horizontal lines. }
    \label{fig:rotoc}
\end{figure}
\textit{Few-body examples.}-- We start with the \textit{kicked rotor model}, a well-studied Floquet chaotic system; see Refs.~\cite{flosprarc12,bitterprl17,bitterpra17} for recent experimental realizations. 
It is defined by the time-dependent Hamiltonian:
\begin{equation}
    H(t) = \frac{1}2 p^2 + K \cos(x) \sum_{n\in \Z} \delta(t - n) \,,
\end{equation}
where $K > 0$ is the kicking strength. Classically, the evolution over a period is given by the standard map:
\begin{equation}
   (x,p) \mapsto \left( x + p ,  p + K \cos(x) \right)   \,. \label{eq:standardmap}
\end{equation}
Ref.~\cite{galitski} studied the classical and quantum OTOC of this model, and found that $\lOTOC > 2\lclas$ for any $K$, with the most pronounced difference occurring in the regime $K \lesssim 1$, where the model is not classically chaotic ($\lclas \approx 0$). We show here that, in that regime, $\lOTOC$ is dominated by the fixed point $(x,p) = (0, 0)$, which has a single unstable exponent:
\begin{equation}
    \omega(K) = \log \left(1+\frac{K}2+\sqrt{K^2 + \frac{K}4} \right) \,. \label{eq:omega_rotor}
\end{equation}
Note that, a fixed point of \eqref{eq:standardmap} corresponds to a periodic orbit, and $\omega(K)$ is the rate at which nearby trajectories deviate from it. Then, it is not hard to adapt the bound \eqref{eq:bound_ex} to the following:
\begin{equation}
    \lOTOC \ge \omega(K) \,, \label{eq:bound_Floquet}
\end{equation} 
which we expect to be tight in the non-chaotic regime. To verify that, we computed the quantum OTOC following the definition and method of Ref.~\cite{galitski}. The results, plotted in Fig.~\ref{fig:rotoc}, show an excellent agreement between $\omega(K)$ and $\lOTOC$ when $K\lesssim 1$. As $K$ further increases, the bound \eqref{eq:bound_Floquet} becomes less tight; at strong couplings ($K \gtrsim 5.4$), the OTOC will be dominated by typical trajectories instead of the saddle. 

\begin{figure}
    \centering
    \includegraphics[width = .8\columnwidth]{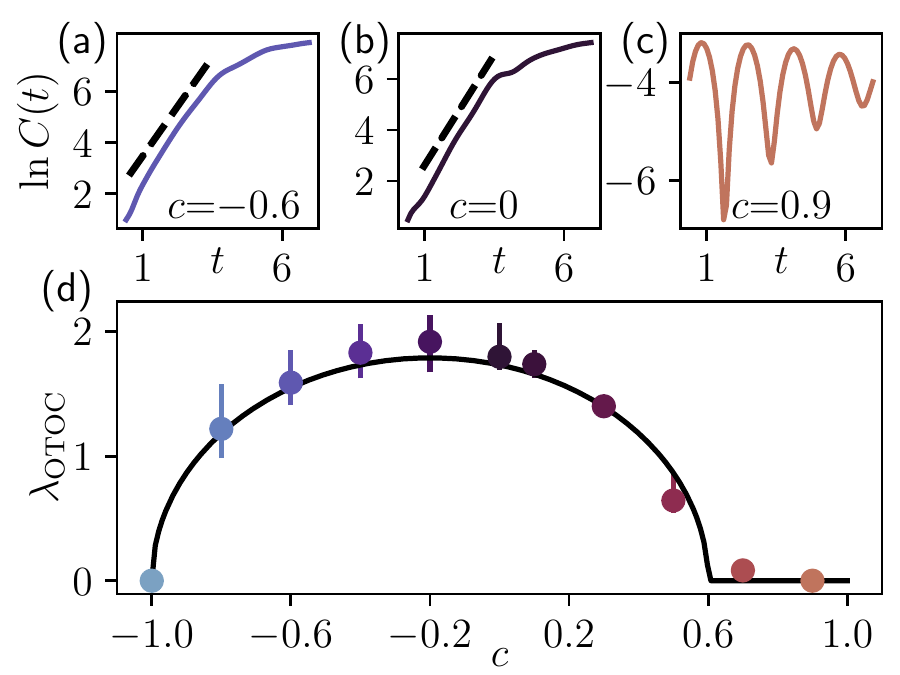}
    \caption{(a-c) Growth of OTOC \eqref{eq:OTOC} where $\hat{O} = \hat{x}_1 + \hat{x}_2$, in the Feingold-Peres model quantized to $S = 75$, and $c=-0.6, 0,$ and $0.9$. The dashed lines are straight lines with slope given by $\omega(c)$ from Eq. \eqref{eq:lowerbound}; for $c = 0.9$, $\omega(c) = 0$, and the OTOC is oscillatory. 
    (d) The data points represent the exponent $\lOTOC$ extracted from the growth of $C(t)$. The continuous curve is $\omega(c)$ from Eq. \eqref{eq:lowerbound}. }
    \label{fig:FP}
\end{figure}
To show that scrambling can be dominated by saddles even in presence of chaos, we consider the \textit{Feingold-Peres (FP)} model of coupled tops, a well-studied few-body chaotic spin model~\cite{feingold83,feingold84pra,fan2017quantum}. Its classical Hamiltonian is 
\begin{equation}
    H = (1+c) (x_1 + x_2) + 4(1-c) z_1 z_2
\end{equation}
where $(x_i, y_i, z_i)$ for $i=1,2,$ are two independent $SU(2)$ spins, and $c\in [-1,1]$ is a parameter. The model is integrable when $c = \pm 1$, and maximally chaotic when $c$ is near $0$ (in the sense of saturating the bound of Ref.~\cite{pcasa18}).
There are no saddles for $c \geq 3/5,$ whereas there are two of them for $c \in [-1, 3/5]$, located at $x_1 = x_2 = \pm 1$, each with one unstable exponent $\omega(c)$.
This leads to the following lower bound:

\begin{equation}
    \lOTOC \ge \omega(c) = \sqrt{(1+c)(3-5c)}  \,,\, -1 \le c \le 3/5  \label{eq:lowerbound}
\end{equation}
and $\omega(c) = 0$ otherwise. To test the tightness of this bound, we computed an OTOC in the quantized FP model, up to $S = 75$ (Hilbert space dimension $\sim 10^4$). In Fig.~\ref{fig:FP}, the extracted $\lOTOC$'s are compared to $\omega(c)$. Surprisingly, the bound~\eqref{eq:lowerbound} turns out to be tight (within error bars) throughout $c\in[-1,1]$: the FP model has saddle-dominated scrambling despite being chaotic. 

A further example of saddle-dominated scrambling, which we delegate to the Supplemental Material, is the Dicke model, well known in atomic physics~\cite{dicke0,dicke,pilatowsky2019positive,rey}.   

\begin{figure}
    \centering
    \includegraphics[width=.8\columnwidth]{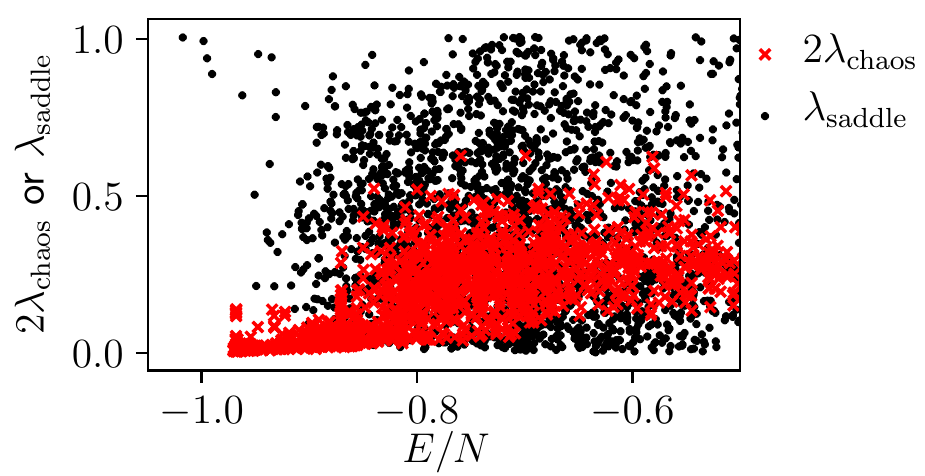}
    \caption{Exponents of the saddle point ($\lsaddle$) \textit{vs} chaos ($2 \lclas$) contributions to the OTOC in the mean-field depinning model~\eqref{eq:Fisher} ($\sigma = 2$ and $N = 128$). In each case, we cool down the system from $T = 2$ gradually to $T= 0.05$, generating along the way $10000$ configurations at different energy densities. To estimate the chaos contribution, we compute the sensitivity up to $t = 50$ starting from each configuration and extract $\lclas$. We plot the resulting $(E/N, 2\lclas)$ as red crosses. To estimate the saddle contributions, we perform gradient descent from each configuration to reach an equilibrium, and compute its $\lsaddle$. The positive values are plotted as dark dots. At low energies, $\lclas$ is severely suppressed, while saddles with large contribution to OTOC are still abundant. } 
    \label{fig:depin}
\end{figure}
\textit{Many-body example.}-- The phenomenon of saddle-dominated scrambling also occurs in many-body systems. A simple example where saddle points naturally occur is provided by the mean-field model of elastic manifolds pinned in a random medium, described by the Hamiltonian~\cite{fisherprl,*fisherPRB}
\begin{equation}
    H = \sum_{j=1}^N \left[\frac12 p_j^2 + V_j(q_j) \right] + \sum_{i,j=1}^N \frac{(q_i - q_j)^2}{2(N-1)}  \,, \label{eq:Fisher}
\end{equation}
where $q_1, \dots, q_N, p_1, \dots, p_N$ are positions and momenta of $N$ degrees of freedom, which interact via an ``all-to-all'' elastic force, while each being pinned in a random potential $V_j$. A convenient choice for the latter is $V_j(q) = \sigma \cos(q + \beta_j)$ where $\beta_j$'s are uniformly distributed in $[0, 2\pi]$ and $\sigma > 0$ is the disorder strength. In the strong disorder regime, such a system is known to have a  complex ``glassy'' energy landscape, with an exponentially large number of equilibria with a wide range of energies~\cite{FYODOROV20181,fyo19,ros19,ledou}. 
Numerically (see caption of Fig.~\ref{fig:depin} for methods), we found a large number of saddle points which have one or few unstable exponents~\footnote{In a different setting where the model is driven, these unstable directions are understood as the precursor of avalanches~\cite{softmodes}}, and for which $\lsaddle$ is positive. In fact, the largest $\lsaddle$'s from low-energy saddles far exceed the typical Lyapunov exponent $\lclas$ at comparable energy, see \fig{depin}. Therefore, scrambling is likely dominated by saddles rather than chaotic trajectories in this model, consistently with our expectations for glassy dynamics: the system is most often trapped around one of an exponential number of local minima; further phase space mixing is achieved by rare crossing of energy barriers, which is the easiest through the vicinity of a saddle point. Nonetheless, we caution that quenched disorder does not guarantee saddle-dominated scrambling: counter-examples include the classical limit of Sachdev-Ye-Kitaev model~\cite{scaffidiSYK}, and the atom-cavity model studied in Ref.~\cite{cavity_paper}.

\textit{Discussion.}-- We have shown that independently of classical chaos, unstable fixed points provide a general mechanism by which out-of-time order correlators (OTOCs) can grow exponentially for an extended period in semiclassical systems. This mechanism turns out to be relevant in several few-body models considered in the recent literature, and can be so in many-body systems as well. Our case studies are by no means exhaustive. In particular, an interesting question is which many-body integrable systems have saddle-dominated scrambling.  


However, our examples make it sufficiently clear that the notion of scrambling, i.e. the exponential growth of OTOCs, is distinct from that of chaos, at least in the semiclassical context. Consequently, the bounds on $\lOTOC$ in Refs.~\cite{pcasa18,Maldacena:2015waa,Avdoshkin,bound_srednicki}, when applied to semiclassical systems, are not only bounds on chaos, but also constrain the instabilities of fixed points and periodic orbits. In particular, this realization makes the bound of Ref.~\cite{pcasa18} on $\lOTOC$ non-trivial even for classical integrable systems.
Distinguishing scrambling from chaos may also affect applications of the former, such as teleportation through a traversable wormhole: for example, the classical protocol of Ref.~\cite{Maldacena:2017axo} (see also~\cite{Gao:2016bin,Gao:2019nyj,Maldacena:2018lmt}) can be realized independently of chaos. Finally, the question remains whether the distinction between chaos and scrambling established here in the semiclassical limit might have an equivalent in the case of strongly coupled quantum systems which have a semiclassical holographic dual. 

\textit{Acknowledgements.}--
It is a pleasure to thank Ehud Altman, Giulio Casati, Jorge Hirsch, Laimei Nie, and Daniel Parker for helpful discussions. Some of the quantum calculations have been performed using the QuSpin package\cite{quspin17,quspin19}. 
TX was supported by the US Department of Energy (DOE), Office of Science, Basic Energy Sciences (BES), under Contract No. AC02-05CH11231
within the Ultrafast Materials Science Program (KC2203).  T.S.
acknowledges support from the Emergent Phenomena
in Quantum Systems initiative of the Gordon and Betty Moore Foundation. XC acknowledges support from ERC synergy Grant UQUAM and DOE grant DE-SC001938.

\bibliography{cmp}

\appendix

\section{Saddle-dominated scrambling in Dicke model}\label{sec:dicke}
\begin{figure}
    \centering
    \includegraphics[width=.8\columnwidth]{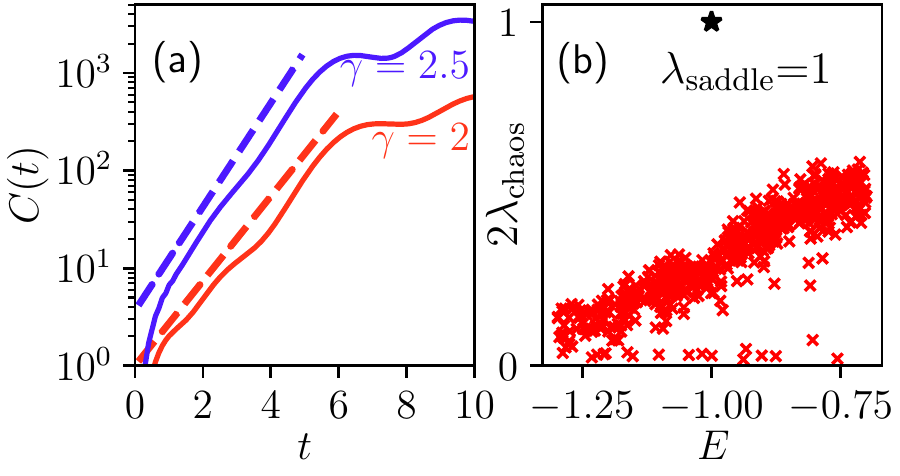}
    \caption{(a) The OTOC in the quantum Dicke model ($N = 40$) with $\hat{O} = \hat{p}$, for a microcanonical ensemble of 40 eigenstates around $H = -1$, and for two representative values of $\gamma$. The dashed slopes show $\lambda_\text{sad}=\omega_1 =  \sqrt{\gamma - 1} $. (b) Classical Lyapunov exponent in the Dicke model with $\gamma=2$ (computed as $(\partial q(t) / \partial q(0))^2 \sim e^{2 \lclas t}$, with $t = 2000$)  for $600$ randomly sampled trajectories in the energy shell $[-1.3, 0.7]$. For all of them, $\lclas$ is smaller than $\lsaddle=\omega_1 = 1$, marked by a star. }
    \label{fig:dicke}
\end{figure}
In this appendix, we show that saddle-dominated scrambling also takes place in the \textit{Dicke model}~\cite{dicke0,dickereview}. Originally, it was conceived to describe $N$ two-level atoms coupled to a single optical cavity mode. Several recent works~\cite{dicke,pilatowsky2019positive,rey} considered the OTOCs in this model. In the classical limit, its Hamiltonian \begin{equation}
    H = \frac12 (q^2 + p^2) + x + \gamma z q 
\end{equation}
describes an $SU(2)$ (pseudo)-spin ($x, y, z$) and a harmonic oscillator ($p, q$), interacting with a coupling constant $\gamma > 0$. A quantum phase transition occurs at $\gamma = 1$, and the super-radiant $\gamma > 1$ phase is characterized by a degenerate pair of ground states, and a saddle at $(q=p=0, x=1)$ with $H=-1$, associated with an excited state transition~\cite{wang19}. The saddle has a single unstable exponent $\omega_1 = \sqrt{\gamma - 1}$, which turns out to be larger than the Lyapunov exponent of a typical classical trajectory with comparable energy, see Fig.~\ref{fig:dicke}(b) and also Refs.~\cite{dicke,pilatowsky2019positive}. This suggests that $\lOTOC \ge \omega_1$ is a tight bound, if the OTOC is computed in an ensemble that overlaps with the saddle point. Our numerics, following the method of~\cite{dicke,pilatowsky2019positive}, confirms this prediction, see Fig.~\ref{fig:dicke}(a).

\end{document}